\documentclass[12pt,thmsa]{article}
\usepackage{eurosym}
\usepackage[left=2.cm, right=2.cm, top=2.2cm, bottom=2.cm]{geometry}
\usepackage{amsfonts}
\usepackage{amsmath}
\usepackage{amssymb}
\usepackage{graphicx}
\usepackage{booktabs}
\usepackage{longtable}
\usepackage{array}
\usepackage{multirow}
\usepackage{wrapfig}
\usepackage{float}
\usepackage{colortbl}
\usepackage{multirow}
\usepackage{pdflscape}
\usepackage{threeparttable}
\usepackage{threeparttablex}
\usepackage[normalem]{ulem}
\usepackage{makecell}
\usepackage{xcolor}
\usepackage{natbib}
\usepackage[nameinlink,noabbrev]{cleveref}
\usepackage{appendix}
\usepackage{authblk}
\usepackage{titling}

\thanksmarkseries{alph}

\begin{document}

\title{Optimal probabilistic forecasts for risk management\thanks{
{\footnotesize All four authors are affiliated with the Department of Econometrics and Business Statistics, Monash University, VIC 3800, Australia. Sun has been support by the Monash Business School Graduate Research Scholarship (from February 2023) and the University of Melbourne Faculty of Business and Economics Graduate Research Scholarship (July 2022 to January 2023); Maneesoonthorn and Martin have been supported by Australian Research Council
(ARC) Discovery Grant DP200101414; and Loaiza-Maya has been supported by ARC
Early Career Researcher Award DE230100029.}} }

\author{Yuru Sun, Worapree Maneesoonthorn\thanks{Corresponding author; Email: Ole.Maneesoonthorn@monash.edu}, Rub\'{e}n Loaiza-Maya and Gael M. Martin}
\maketitle

\begin{abstract}
This paper explores the implications of producing forecast distributions that are optimized according to scoring rules that are relevant to financial risk management. We assess the predictive performance of optimal forecasts from potentially misspecified models for i) value-at-risk and expected shortfall predictions; and ii) prediction of the VIX volatility index for use in hedging strategies involving VIX futures. Our empirical results show that calibrating the predictive distribution using a score that rewards the accurate prediction of extreme returns improves the VaR and ES predictions. Tail-focused predictive distributions are also shown to yield better outcomes in hedging strategies using VIX futures.\bigskip

 \bigskip

\emph{MSC2010 Subject Classification}: 60G25, 62M20, 91G70\smallskip

\end{abstract}

\newpage

\baselineskip 18pt

\section{Introduction}

\label{sec:intro}

Probabilistic forecasts -- produced via full predictive probability
distributions -- provide complete information about the future values of a
random variable. Conventional approaches to probabilistic prediction have
assumed: either that the predictive model (or likelihood function) correctly
specifies the true data generating process (DGP), or that some specified set
of predictive models spans the truth; neither of which is likely to be the
case in practice.\footnote{%
We make it clear from the outset that we use the terms `forecast' and
`prediction' (and all of their various grammatical derivations) synonymously
and interchangeably, for the sake of linguistic variety. We note that the
term `prediction' is a broader term, also used for non-temporal settings,
for example.}

\textit{Optimal }probabilistic forecasts, on the other hand, are based on
predictive distributions that are optimal in some user-specified proper
scoring rule (\citealp{gneiting2005calibrated}; \citealp{gneiting2007}; %
\citealp{patton2019comparing}; \citealp{optimal}; Zischke et al, 2022). That
is, instead of defaulting to the conventional use of a (misspecified)
likelihood function, such forecasts exploit a criterion function built from
a scoring rule that rewards the type of predictive accuracy that matters for
the problem at hand. Optimization may occur with respect to: the parameters
of a given model, the weights of a combination of predictive models, or
both. In the case of weighted combinations, the approach forms part of a
broader push -- based on both Bayesian and frequentist principles -- to
construct forecast combinations that are explicitly designed to perform well
according to a particular forecasting metric (see \citealp{fbp2020} for a
recent contribution, and relevant coverage of the literature; and %
\citealp{Aastveit2019} and \citealp{Wang2022} for more extensive reviews).

In short, optimal probabilistic forecasts aim to produce predictions that
are accurate in the way that \textit{matters}, irrespective of the
specification of the predictive model (or set of predictive models). In the
context of financial risk management, for example, the tails of the
predictive distribution of a financial return are of particular interest,
with the lower tail capturing the risk of loss in a long (buying) position,
while the upper tail captures the risk of loss in a short-selling position.
To date, the literature on risk prediction has focused on the development of
alternative models that better capture the tail features of returns via:
alternative specifications for volatility (e.g. \citealp{ding1993}, and %
\citealp{hansen2012}), alternative conditional distributions for the return
(e.g. \citealp{chen2008}, and \citealp{bensaida2018}), the addition of more
latent factors (\citealp{maheu2004}, and \citealp{2maneesoonthorn2017}), or
specifications that model only specific quantiles \citep{engle2004caviar} or
specific tail regions \citep{brooks2005}. Scoring rules that are relevant to
the tails may then be used to \textit{evaluate} the forecasts, but do not
play a role in the construction of the forecasts themselves (see %
\citealp{giacomini2005}, \citealp{jensen2013}, and \citealp{chiang2021},
amongst many others, including those already referenced above).

In this paper, we take a difference stance. We adopt a predictive model that
is \textit{known} to miss certain critical features of a financial return --
namely a generalized autoregressive conditional heteroscedastic (GARCH)
model with Gaussian innovations \citep{bollerslev} -- but which is simple
and quick to estimate. We then calibrate the model using proper scoring
rules that are relevant to the tail region in question, and construct an
optimal predictive accordingly. We show how calibrating a Gaussian
GARCH(1,1) model to tail-focused scoring rules can yield accurate
predictions of extreme returns, and illustrate how the results vary with the
degree of model misspecification. We also assess the ability of the optimal
predictives to produce accurate predictions of value-at-risk (VaR) and
expected shortfall (ES) predictions, and compare that accuracy with
likelihood-based (or, equivalently, logarithmic score (LS)-based)
probabilistic forecasts. We employ two specific tail-focused rules for the
construction of optimal forecasts: the censored likelihood score (CLS) of %
\citet{diks2011} and the quantile score (QS) (\citealp{giacomini2005}, and %
\citealp{gneiting2007}). The CLS is a proper scoring rule that allows a
forecaster to focus on accuracy in any particular region of interest,
including the tail of the predictive distribution. As a proper scoring rule
for quantiles, the QS is also particularly relevant here for the production
of VaR predictions, having also been used previously as the criterion
function in the estimation of quantile regressions (\citealp{koenker1978}
and \citealp{engle2004caviar}). The VaR and ES predictions are assessed
using the conditional coverage test of \citet{christoffersen1998} and the
recently proposed ES predictive dominance test of \citet{ziegel2020robust},
respectively.

We then extend our application of optimal forecasts to the problem of
predicting the option-implied volatility index (VIX), and producing
effective hedging strategies based on VIX derivatives (\citealp{moran2020vix}%
; \citealp{konstantinidi2008}; \citealp{taylor2019}). Once again, we depart
from the idea of seeking an ideal predictive model for the index (%
\citealp{konstantinidi2008}; \citealp{fernandes2014}; \citealp{taylor2019}),
instead constructing optimal forecasts using the simple heterogenous
autoregressive (HAR) model of \citet{corsi2009} with Gaussian GARCH(1,1)
errors. In this case we use the CLS score to calibrate the model -- and
produce the optimal predictive -- but with the region of `focus' designed to
match certain features of the trading strategy in question.

The paper proceeds as follows. Section \ref{sec:score} explains how scoring
rules are used to construct optimal\ distributional forecasts, as well as
outlining the specific tail-focused rules that are relevant to financial
risk management. Section \ref{sec:sim} investigates the benefits yielded by
optimal forecasting in a simulation setting, paying particular attention to
how well the optimal forecasts perform in predicting tail risks via the
assessment of VaR and ES predictions for a financial return. Model
misspecification is explicitly factored into the simulation design with a
view to replicating a realistic empirical scenario. An extensive empirical
analysis of 12 S\&P market and industry indices is then conducted in
Section \ref{sec:returns}, with the results confirming our simulation
findings in Section \ref{sec:sim}: i.e. that the optimal predictives
achieve substantially better tail risk predictions than the likelihood-based alternatives. In Section \ref{sec:vix}, we explore the use of
optimal forecasts of the VIX index in the construction of effective hedging
strategies using recent data on the VIX index and VIX futures. The ability
of the tailored score-based predictives to achieve better investment
outcomes highlights the practical importance of optimal forecasting in risk
management. We conclude in Section \ref{sec:conclude}.

\section{Scoring Rules in Risk Prediction}

\label{sec:score}

\subsection{Scoring Rules and the Optimal Forecast Distribution}

Scoring rules \citep{gneiting2007} are measures of performance for
probabilistic forecasts. As such, they are often used to evaluate and rank
competing probabilistic forecasts, based on materialized events or values,
by assigning a numerical score. That is, if a forecaster produces a
predictive distribution $\mathit{P}$ and the event $y$ materializes, then
the reward for this prediction given by a scoring rule is denoted by $S(P,y)$%
. For a positively-oriented score, a higher value will be assigned to the
better of two competing predictions, on the condition of the scoring rule
being proper. The propriety of the scoring rule is crucial here, as an
improper scoring rule may assign a higher average score to an incorrect
prediction, i.e. one that does not tally with the true DGP.

More formally, suppose $P$ is the predictive distribution and $Q_{0} $ is
the true DGP. $\mathbb{S}(P,Q_{0})$ denotes the expected value of $S(P,\cdot
)$ under $Q_{0}$. That is, 
\begin{equation}
\mathbb{S}(P,Q_{0})=\int_{y\in \Omega }S(P,y)dQ_{0}(y)  \label{eq:score}
\end{equation}%
where $\Omega $ is the sample space. A scoring rule is said to be proper if $%
\mathbb{S}(Q_{0},Q_{0})\geq \mathbb{S}(P,Q_{0})$ for all $P$ and $Q_{0}$,
and is strictly proper if $\mathbb{S}(Q_{0},Q_{0})=\mathbb{S}(P,Q_{0})$ only
when $P=Q_{0}$ \citep{gneiting2007}. This ensures that a proper scoring rule
is maximized when the prediction reveals the truth, on the condition that
the predictive distribution class $\mathcal{P}$ contains $Q_{0}$.

In practice of course, the expected score $\mathbb{S}(\cdot ,Q_{0})$ cannot
be obtained, but we can use a sample average of observed scores as a
reasonable estimator. Asymptotically, the true predictive distribution can
be recovered by optimizing a sample criterion based on any proper scoring
rule, if the true predictive is contained in the predictive class over which
the maximization occurs. Even when $Q_{0}$ is not contained in the
predictive distribution class $\mathcal{P}$ proposed by the forecaster, it
does not change the fact that a proper scoring rule will reward a particular
form of forecast accuracy and, hence, optimization will select the best
forecast within the predictive class, according to this scoring rule.

Thus, a scoring rule can be used to define a criterion function that can be
optimized to obtain parameter estimates, from which a probabilistic
prediction that is optimal in that score can be produced. Assume that the
scoring rule is positively-oriented and the unknown parameters of the
assumed predictive model are denoted as $\boldsymbol{\theta }\in \Theta $,
and let $P_{\boldsymbol{\theta }}^{t-1}$ := $P(\cdot |\mathcal{F}_{t-1},%
\boldsymbol{\theta })$ be the one-step-ahead predictive distribution
function based on the assumed model, where $\mathcal{F}_{t-1}$ represents
all the information available at time $t-1$, and $p(\cdot |\mathcal{F}_{t-1},%
\boldsymbol{\theta })$ be the corresponding predictive density function at
time $t$. For $\tau $ such that $T\geq \tau \geq 1$, let $%
\{y_{t}\}_{t=1}^{T-\tau }$ denote a series of size\textbf{\ }$T-\tau $ used
for the optimization, with $\tau $ denoting the size of a {hold-out sample}.
Defining:%
\begin{equation}
\bar{S}(\boldsymbol{\theta })=\frac{1}{T-(\tau +1)}\sum_{t=2}^{T-\tau }S(P_{%
\boldsymbol{\theta }}^{t-1},y_{t}),  \label{eq:sbar}
\end{equation}%
then an estimator $\boldsymbol{\hat{\theta}}$ obtained by maximizing $\bar{S}%
(\boldsymbol{\theta })$:%
\begin{equation}
\boldsymbol{\hat{\theta}}=\text{arg}\max_{\boldsymbol{\theta }\in \Theta }%
\bar{S}(\boldsymbol{\theta })  \label{eq:1}
\end{equation}%
is said to be \textquotedblleft optimal\textquotedblright\ based on this
scoring rule, and $P_{\boldsymbol{\hat{\theta}}}^{t-1}$ is referred to as
the optimal predictive. Under certain conditions, including that the scoring
rule is proper and that the predictive model is correctly specified, $%
\boldsymbol{\hat{\theta}}\rightarrow \boldsymbol{\theta }_{0}\ \text{as}\ 
\mathit{T}\rightarrow \infty $ (for a fixed $\tau $), where $\boldsymbol{%
\theta }_{0}$ represent the true (vector) parameter. If the predictive model
is misspecified, then $\boldsymbol{\hat{\theta}}$ converges to $\boldsymbol{%
\theta _{\ast }}$, where the latter denotes the maximum of the limiting
criterion function to which $\bar{S}(\boldsymbol{\theta })$ converges as $T$
diverges. (\citealp{gneiting2007}; \citealp{optimal}; Zischke et al, 2022.)

\subsection{Scoring Rules for Financial Risk Predictions\label{scores_pred}}

In risk management applications, the region of interest is typically in the
tails of the distribution, with the lower tail being relevant to the risk of
loss in long positions and the upper tail relevant to the risk of loss in
short positions. The censored logarithmic score (CLS) of \cite{diks2011}
allows users to reward accurate forecasts in a particular region (or
regions) of interest, such as an upper or lower tail. It is defined as 
\begin{equation}
S_{CLS}(P_{\boldsymbol{\theta }}^{t-1},y_{t})=I(y_{t}\in A)\log \left(
p(y_{t}|\mathcal{F}_{t-1},\boldsymbol{\theta })\right) +I(y_{t}\in
A^{c})\log \left( \int_{y\in A^{c}}p(y|\mathcal{F}_{t-1},\boldsymbol{\theta }%
)dy\right) ,  \label{eq:cls}
\end{equation}%
where $A$ is the region of interest and $A^{c}$ is its complement. The CLS
has been applied in the evaluation of predictive distributions geared to
risk management (\citealp{jensen2013}, and \citealp{chiang2021}), as well as
in inferential procedures geared to certain data regions (%
\citealp{gatarek2014}, and \citealp{borowska2020}). \citet{opschoor2017}
have also applied the CLS in the estimation of forecast combinations and
found that density forecasts based on optimizing the CLS outperform
alternatives forecasts that are not constructed to reward tail accuracy (see
also \citealp{fbp2020}).

A feature of the predictive distribution that is critically important in
financial risk management is the VaR: computed (in the case of one-step-ahead prediction) as the $p\times 100\%$ quantile of the one-step-ahead
predictive distribution ${P_{\boldsymbol{\theta }}^{t-1}}$, and denoted by $%
VaR_{p}(P_{\boldsymbol{\theta }}^{t-1})$. The quantile score (QS) is a
proper score for quantiles and is thus particularly suitable for rewarding
accurate prediction of the VaR (see, for example, \citealp{giacomini2005}, %
\citealp{bao2006} and \citealp{laporta2018}). The QS is defined, in turn, as 
\begin{equation}
S_{QS}(P_{\boldsymbol{\theta }}^{t-1},y_{t})=\left[ I(y_{t}\leq VaR_{p}(P_{%
\boldsymbol{\theta }}^{t-1}))-p\right] \left( y_{t}-VaR_{p}(P_{\boldsymbol{%
\theta }}^{t-1})\right) .  \label{eq:qs}
\end{equation}

In this paper, we investigate whether calibrating the model to the
tail-focused proper scoring rules, such as (the appropriately specified) CLS
and QS, yields accurate probabilistic forecasts for financial risk
management. As a comparator we adopt the LS, defined as

\begin{equation}
S_{LS}(P_{\boldsymbol{\theta }}^{t-1},y_{t})=\log \left( p(y_{t}|\mathcal{F}%
_{t-1},\boldsymbol{\theta })\right) ,
\end{equation}%
which, as a `local' scoring rule, will assign a high value if the realized
value of $y_{t}$ is in the high density region of $p\left( y_{t}|\mathcal{F}%
_{t-1},\boldsymbol{\theta }\right) $ and a low value otherwise. A sample
criterion defined using the log score is equivalent to the log-likelihood
function. Thus, optimization with respect to the LS yields the maximum
likelihood estimator of $\boldsymbol{\theta }$ as%
\begin{equation}
\boldsymbol{\hat{\theta}}_{MLE}=\text{arg}\max_{\boldsymbol{\theta }\in
\Theta }\left\{ \frac{1}{T-(\tau +1)}\sum_{t=2}^{T-\tau }\log \left( p(y_{t}|%
\mathcal{F}_{t-1},\boldsymbol{\theta })\right) \right\} ,
\end{equation}%
which, under the assumption of correct specification (and regularity), is
the asymptotically efficient estimator of $\boldsymbol{\theta }_{0}$; hence
the interest in documenting the performance of what may be viewed as the
conventional approach to the construction of a predictive density.

Finally, we note that the expected shortfall (ES) is a point forecast that
is often reported alongside the VaR to evaluate the tail risk in financial
investments, and is considered to be a robust risk measurement, given its
coherence properties (see \citealp{acerbi2002coherence}, %
\citealp{tasche2002expected}, and \citealp{yamai2005value}). By definition,
the $p^{th}$ ES constructed from the one-step-ahead predictive distribution $%
P_{\boldsymbol{\theta }}^{t-1}$ is defined as 
\begin{equation}
ES_{p}(P_{\boldsymbol{\theta }}^{t-1})=\frac{1}{p}\int_{0}^{p}VaR_{\alpha
}(P_{\boldsymbol{\theta }}^{t-1})d\alpha ,  \label{es}
\end{equation}%
where $VaR_{\alpha }(P_{\boldsymbol{\theta }}^{t-1})$ denotes the $\alpha
\times 100\%$ quantile of the predictive distribution. The ES is not
elicitable as a point forecast, implying that there is no scoring function
that allows for a consistent ranking of model performance based on ES alone %
\citep{3gneiting2011}. However, \citet{fissler2015expected} propose a class
of scoring functions that can \textit{jointly} elicit the forecast accuracy
of VaR and ES, and can thus be used to measure ES predictive accuracy (in
conjunction with that of VaR). The joint scoring function for $VaR_{p
}(P_{\boldsymbol{\theta }}^{t-1})$ and $ES_{p}(P_{\boldsymbol{\theta }%
}^{t-1})$, in positive orientation, is defined as 
\begin{equation}
\begin{aligned} S_\eta(P^{t-1}_{\boldsymbol{\hat \theta}},y_t) =-I(\eta \le
ES_p(P^{t-1}_{\boldsymbol{\hat \theta}}))&\left(\frac{1}{p}I(y_t \le
VaR_{p}(P^{t-1}_{\boldsymbol{\hat
\theta}}))(VaR_{p}(P^{t-1}_{\boldsymbol{\hat \theta}})-y_t) \right. \\
&-\left. (VaR_{p}(P^{t-1}_{\boldsymbol{\hat \theta}})-\eta) \right) -I(\eta
\le y_t)(y_t - \eta). \end{aligned}  \label{eq:escore}
\end{equation}%
Here, the scoring function is indexed by the real number $\eta $ that spans
the range of the random variable of interest, with the \textit{dominance }of
one predictive model over another (in terms of ES accuracy) gauged by values
of their respective scores over a relevant range of $\eta $ (see %
\citealp{ehm2016quantiles}, and \citealp{ziegel2020robust}). Whilst we do
not use this scoring function to determine the criterion used to produce an
estimate of $\boldsymbol{\theta }$ and, ultimately, the corresponding
optimal predictive, we do adopt this scoring function to compare the
accuracy of the ES (and VaR) predictions produced by the relevant
tail-focused predictives, with the accuracy of those produced by the LS- (or
MLE-based) predictive.

\section{Performance of Score-Based Risk Predictions:\ Simulation Evidence}

\label{sec:sim}

Forecasts explicitly designed to reward certain types of accuracy -- whether
it be via the optimization advocated in this paper, or via a suitable
(``focused'') Bayesian posterior up-date -- can certainly outperform
predictions produced by conventional methods; see \citet{opschoor2017}, %
\citet{fbp2020}, and \citet{optimal}. In simulation settings it has been
shown that the degree of model misspecification has a notable effect on the
performance of optimal forecasts. In short, the greater the degree of
misspecification, the greater the gain from optimal (or focused) forecasts,
at least conditional on the assumed predictive model being broadly
``compatible''\ with the true DGP.\footnote{%
We refer readers to \citet{optimal} for discussion of this point.} This
result augurs well for the usefulness of such targeted predictions in
empirical settings, in which misspecification is unavoidable, and this
provides strong motivation for the analysis undertaken herein -- in which
optimal predictions are sought in empirical risk management settings. In
this section, we evaluate the performance of the optimal tail-focused
forecasts, relative to the conventional MLE-based forecast, in terms of
their ability to predict tail risks in a simulation setting. Whilst the
MLE-based predictive is formally ``optimal'' with respect to the LS, we
reserve the term ``optimal'' hereafter for the predictives based on the
tail-focused scores.

\subsection{Simulation Design}

\label{sec:sim_design}

We generate $T$ observations of a random variable $y_{t}$ from three
alternative true DGPs, each of which captures
the main stylized features of a financial return and its volatility. A
GARCH model of
order (1,1) with Gaussian errors is used to define the predictive class $%
\mathcal{P}$ in all experiments. Scenario (i) represents the case of correct
specification, with the true DGP matching the predictive model is this case.
Scenario (ii) uses a GARCH(1,1) model with Student-t errors as the true DGP,
thereby representing the case of mispecification of the fourth moment.
Finally, in Scenario (iii) the skewed stochastic volatility model
introduced by \citet{smith2018}, and also used in \citet{fbp2020}, is used
as the true DGP. This model assumes a skew-normal marginal distribution,
which allows us to measure the performance of the optimal forecasts under
misspecification of the third moment of the data, in addition to the use of
an incorrect conditionally deterministic volatility specification. Under
each of Scenarios (ii) and (iii), we vary the degree of misspecification by
changing the degrees of freedom in the Student-t distribution and the shape
parameter in the skew-normal distribution, respectively. Details of each of
these scenarios are given in Table \ref{tab:sim_design}.

\begin{table}[h]
\caption{The three simulation scenarios used in the numerical analysis. The
assumed predictive model is GARCH(1,1) in all cases, and the true DGP is
varied across the scenarios. Scenario (i) represents correct model
specification. Scenario (ii) represents the first form of misspecification,
where $t_{\protect\nu }$ indicates a Student-t distribution, with the
degrees of freedom parameter, $\protect\nu $, assuming values of 3 and 12.
Scenario (iii) represents the second form of misspecification, with the
shape parameter in the skew-normal marginal return distribution assuming the
values of 0 (representing a symmetric marginal distribution), and -3 and -5
(which produce, in turn, increasing negative skewness in the marginal
distribution.}
\label{tab:sim_design}\centering%
\resizebox{\textwidth}{!}{
\begin{tabular}{cccccc}
\hline
& & & &&\\
          & \textbf{Scenario (i)} && \textbf{Scenario (ii)} && \textbf{Scenario (iii)} \\
                    & & & &&\\
\cline{2-2}\cline{4-4}\cline{6-6}
          & & &&& \\
\multirow{4}{*}{\textbf{True DGP}} & $y_t = \sigma_t \epsilon_t$ & &$y_t = \sigma_t \epsilon_t$ & & $y_t=D^{-1}(F_z(z_t))$ \\
& $\sigma_t^2 = 1 + 0.2y_{t-1}^2+0.7\sigma^2_{t-1}$ && $\sigma_t^2 = 1 + 0.2y_{t-1}^2+0.7\sigma_{t-1}^2$ && $z_t = \exp (h_t/2) \epsilon_t$, $\epsilon_t\sim i.i.d.N(0,1)$ \\ 
& $\epsilon_t \sim i.i.d.N(0,1)$ & &$\epsilon_t \sim i.i.d.(\frac{\nu-2}{\nu})^{0.5} * t_\nu$ & &  $h_t=-0.4581+0.9(h_{t-1}+0.4581)+0.4172 \eta_t$, $\eta_t \sim i.i.d.N(0,1)$\\ 
& && $\nu \in \{3,12\}$ & & $D(.)$ is the CDF of a skew-normal distribution\\ 
& & &&&  with shape parameter $\in \{0,-3,-5\}$\\
& & &&&  \\
\multirow{3}{*}{\textbf{Assumed model}} & $y_t = \mu +\sigma_t \epsilon_t$ && $y_t = \mu +\sigma_t \epsilon_t$ && $y_t = \mu +\sigma_t \epsilon_t$ \\ 
& $\sigma_t^2 = \alpha_0 + \alpha_1 (y_{t-1}-\mu)^2+\beta_1\sigma_{t-1}^2$ && $\sigma_t^2 = \alpha_0 + \alpha_1 (y_{t-1}-\mu)^2+\beta_1\sigma_{t-1}^2$ && $\sigma_t^2 = \alpha_0 + \alpha_1 (y_{t-1}-\mu)^2+\beta_1\sigma_{t-1}^2$ \\ 
& $\epsilon_t \sim i.i.d.N(0,1)$ && $\epsilon_t \sim i.i.d.N(0,1)$ && $\epsilon_t \sim i.i.d.N(0,1)$ \\
& & & &&\\
\hline
\end{tabular}}
\end{table}

We estimate optimal forecasts under all three scenarios using the two proper
scoring rules introduced in Section \ref{scores_pred}: the CLS and the QS.
For the CLS, we focus on the 10\% lower tail, corresponding to the risk
associated with a long financial position, and we label this as CLS10
hereafter. In order to mimic the situation that prevails in practice, in
which the true predictive is unknown, the region $A$, required for
evaluation of the CLS, is based on a marginal empirical quantile rather than
a conditional quantile. The QS is used as the scoring rule that targets the
predictive accuracy of a $p\times 100\%$ conditional quantile. We consider
quantiles that are relevant to the computation of VaR and ES quantities in
practice, by specifying $p\times 100\%=\{2.5\%,5\%,10\%\}$. We label these
three versions of QS as QS2.5, QS5 and QS10 respectively. Because the
LS-based predictive gives us insight into the performance of the
conventional MLE method, it is used as a benchmark for assessing the
improvement of the optimal forecasts.

Let $P_{\boldsymbol{\hat{\theta}_{i}}}^{t-1}$ denote the one-step-ahead
predictive distribution function based on the Gaussian GARCH(1,1) model,
optimized according to scoring rule $S_{i}$. For each of the true DGPs
specified in Table \ref{tab:sim_design}, we draw $T=6000$ observations of $%
y_{t}$ from $Q_{0}$. Given $\tau =5000$, we then utilize expanding
estimation windows, beginning with a sample of size 999, to obtain the
optimal one-step-ahead predictive $P_{\boldsymbol{\hat{\theta}_{i}}}^{t-1}$,
for $t=T-\tau +1,...,T$. For each $t$, $\boldsymbol{\hat{\theta}}_{i}=\{\hat{%
\mu},\hat{\alpha}_{0},\hat{\alpha}_{1}\}_{i}$ is obtained as in (\ref{eq:1})
and (\ref{eq:sbar}) for $S_{i}\in \{LS,CSL10,QS2.5,QS5,QS10\}$. We then
document the relative performance of the MLE-based predictive and the
optimal predictives in two ways. First, we assess the performance of all in
producing accurate VaR predictions by conducting the conditional coverage
test of \citet{christoffersen1998}. Forecasts that produce VaR exceedences
that are insignificantly different from the nominal $p$ level, and
that are uncorrelated over time, are deemed to be desirable. Second, we
evaluate the accuracy of the ES forecasts using the scoring rule defined
in (\ref{eq:escore}), along with the Murphy diagram of \cite{ehm2016quantiles} and
\citet{ziegel2020robust}, where, as noted earlier, this diagram is designed to assess the
ES predictive dominance of one predictive over another.

\subsection{Value-at-Risk Assessment}

\label{sec:simvar}

In this section, we\ assess the accuracy of the 2.5\%, 5\% and 10\% VaR
predictions constructed from the predictive densities $P_{\boldsymbol{\hat{%
\theta}_{i}}}^{t-1}$, for $S_{i}\in \{LS,CLS10,QS2.5,QS5,QS10\}$. The
empirical tail coverage and conditional coverage test results are documented
in Table \ref{tab:garch_var}. The bolded values indicate instances where the
null hypothesis of the conditional coverage test is not rejected at the 5\%
significance level. The bolded \textit{italicized} value in a given column
indicates the (insignificant) result with the empirical tail coverage that
is the closest to the nominal coverage.

Panel A of Table \ref{tab:garch_var} reports the results for the correctly
specified case. As anticipated, given that all scoring rules are proper, and
the (expanding) estimation samples are large, optimization of (\ref{eq:sbar}%
) -- for each choice of score -- produces a value for $\boldsymbol{\hat{\theta}%
}$ that is close to the true value of $\boldsymbol{\theta }_{0}$; hence all
predictives are very similar and have comparable out-of-sample performance,
however measured -- and this is what we see (overall) in the Panel A results.

In contrast, if one considers what is arguably the most misspecified case --
Scenario (iii) with shape parameter of -5 -- with results reported in Panel
F, we see evidence of what \citet{optimal} refer to as ``strict coherence''.
That is, the scoring rule designed to yield optimal predictions according to
a particular measure produces the \textit{best} performance out of sample in
terms of that measure. Specifically, the QS($p\times 100$)-based predictive
produces the most accurate prediction of the VaR at that nominal level,
for each value of $p.$ Moreover, the CLS10-based predictive -- which focuses
on lower tail accuracy \textit{per se} -- performs almost as well as each QS($p\times 100$)-based predictive, yielding an empirical
coverage that is insignificantly different from the nominal value for all three VaRs considered. In the case of Scenario (ii), Panel C, in which the
misspecification manifests itself via a very fat-tailed Student-t innovation
that is not matched by the (conditionally) Gaussian predictive model, strict
coherence is also in evidence for the 5\% and 10\% VaRs.

For the remaining misspecified cases, there is certainly some evidence that
the tail-focused scores reap benefits out-of-sample in terms of VaR
prediction. Most notably, the QS-based optimal forecasts \textit{always}
accurately predict the relevant VaR (i.e. the corresponding values recorded in the table are always bolded), independent of the degree of
misspecification; that is, even if the QS-based predictive is not the most
accurate in every \textit{single} case, it never behaves poorly, auguring
well for the blanket use of a QS-based predictive in empirical settings in which accurate VaR
coverage is the goal.

\begin{table}[tbph]
\caption{Predictive performance for VaR under Scenarios (i)-(iii) in Table 
\protect\ref{tab:sim_design}. The numbers reported are the proportion of
exceedances. The bold font is used to indicate exceedances that are not
significantly different from the nominal level and are uncorrelated over
time, resulting in a failure to reject the null hypothesis of the
conditional coverage test, at the 5\% level of significance. The (bolded)
italicized value in a given column is the exceedance that fails to reject
the null hypothesis of the conditional coverage test \textit{and} is closest
to the nominal $p$ level.}
\label{tab:garch_var}
. \centering%
\resizebox{\textwidth}{!}{
\begin{tabular}{lccc|ccc}
 & \multicolumn{6}{c}{}\\ 
\toprule\toprule & \multicolumn{6}{c}{Out-of-sample exceedances} \\ 
\cmidrule{2-7} Optimizers & \multicolumn{1}{l}{VaR at 2.5\%} & 
\multicolumn{1}{l}{VaR at 5\%} & \multicolumn{1}{l|}{VaR at 10\%} & \multicolumn{1}{l}{VaR at 2.5\%} & 
\multicolumn{1}{l}{VaR at 5\%} & \multicolumn{1}{l}{VaR at 10\%}\\ 
\midrule &  &  &  &&& \\ 
& \multicolumn{3}{c|}{\textbf{Panel A: Scenario (i)}}  & \multicolumn{3}{c}{\textbf{Panel D: Scenario (iii) with Shape=0}}\\ 
&  &  &  & & &  \\ 
MLE & \textit{\textbf{2.94\%}} & \textbf{5.24\%} & \textbf{10.18\%}  & \textbf{2.30\%} & \textit{\textbf{5.12\%}} & \textbf{10.82\%}\\ 
CLS10 & 3.26\% & \textbf{5.52\%} & \textbf{9.98\%} & \textbf{2.42\%} & \textit{\textbf{5.12\%}} & 10.46\% \\ 
QS2.5 & \textbf{2.98\%} & \textit{\textbf{4.86\%}} & \textbf{9.36\%}  & \textbf{2.56\%} & \textbf{5.38\%} & 10.76\%\\ 
QS5 & 3.38\% & \textbf{5.70\%} & \textit{\textbf{10.00\%}}  & \textit{\textbf{2.46\%}} & \textbf{5.14\%} & 10.72\%\\ 
QS10 & 3.50\% & 5.90\% & \textbf{10.60\%}  &\textbf{2.26\%} & \textbf{4.66\%} & \textit{\textbf{10.46\%}}\\ 
\midrule &  &  &   & & &\\ 
& \multicolumn{3}{c|}{\textbf{Panel B: Scenario (ii) with $\nu$ = 12 }}  & \multicolumn{3}{c}{\textbf{Panel E: Scenario (iii) with Shape=-3}}\\ 
&  &  &  \\ 
MLE & 2.98\% & \textbf{5.38\%} & \textbf{9.72\%}  & 4.10\% & 7.08\% & 11.76\%\\ 
CLS10 & \textit{\textbf{2.78\%}} & \textbf{5.28\%} & \textit{\textbf{10.00\%}} & \textit{\textbf{2.50\%}} & \textit{\textbf{5.10\%}} & \textbf{9.54\%} \\ 
QS2.5 & \textbf{2.82\%} & \textit{\textbf{4.88\%}} & 9.06\%  & \textbf{2.54\%} & \textbf{4.62\%} & 8.60\%\\ 
QS5 & 3.10\% & \textbf{5.32\%} & \textbf{9.40\%}  & \textbf{2.88\%} & \textbf{5.24\%} & 9.16\% \\ 
QS10 & 3.80\% & 6.06\% & \textbf{10.40\%}  & 3.52\% & 6.06\% & \textit{\textbf{10.38\%}}\\ 
\midrule &  &  &   & & &\\ 
& \multicolumn{3}{c|}{\textbf{Panel C: Scenario (ii) with $\nu$ = 3 }}  &  \multicolumn{3}{c}{\textbf{Panel F: Scenario (iii) with Shape=-5}}\\ 
&  &  &   & & &\\ 
MLE & \textit{\textbf{2.70\%}} & 4.00\% & 7.06\%  &4.54\% & 7.66\% & 12.06\% \\ 
CLS10 & \textbf{2.14\%} & 4.24\% & 10.96\%  & \textbf{2.66\%} & \textbf{5.24\%} & \textbf{9.48\%} \\ 
QS2.5 & \textbf{2.92\%} & 4.32\% & 7.24\%  &\textit{\textbf{2.58\%}} & \textbf{4.70\%} & 8.22\%\\ 
QS5 & 4.10\% & \textit{\textbf{5.28\%}} & 7.96\%  & \textbf{2.80\%} & \textit{\textbf{5.20\%}} & 8.88\%\\ 
QS10 & 6.20\% & 7.78\% & \textit{\textbf{10.14\%}}  &3.72\% & 6.26\% & \textit{\textbf{10.38\%}}\\ 
\bottomrule\bottomrule 
\end{tabular}}
\end{table}

\subsection{Expected Shortfall Assessment}

\label{sec:es_sim} Based on (\ref{eq:escore}), \citet{ziegel2020robust}
propose the concept of \textquotedblleft forecast
dominance\textquotedblright\ for comparative assessment of joint pairs of
VaR and ES. The predictive $P_{\boldsymbol{\hat{\theta}_{i}}}^{t-1} $ is
deemed to weakly dominate the predictive $P_{\boldsymbol{\hat{\theta}_{j}}%
}^{t-1}$, when $E[S_{\eta }(P_{\boldsymbol{\hat{\theta}_{i}}%
}^{t-1},y_{t})]\geq E[S_{\eta }(P_{\boldsymbol{\hat{\theta}_{j}}%
}^{t-1},y_{t})]$ for any value of $\eta $. We follow \citet{ziegel2020robust}
in assessing the dominance of competing predictives using the Murphy
diagram, where the horizontal axis is $\eta $ and the vertical axis is the
average score difference constructed as

\begin{equation}
\Delta _{\eta }\left( P_{\boldsymbol{\hat{\theta}_{i}}}^{t-1},P_{\boldsymbol{%
\hat{\theta}_{MLE}}}^{t-1}\right) =\widehat{E}\left[ S_{\eta }(P_{%
\boldsymbol{\hat{\theta}_{i}}}^{t-1},y_{t})\right] -\widehat{E}\left[
S_{\eta }(P_{\boldsymbol{\hat{\theta}_{MLE}}}^{t-1},y_{t})\right] ,
\label{murph}
\end{equation}%
where the expectations are estimated as the sample mean of the relevant
out-of-sample scores, in the usual way. In order to keep the analysis
manageable, we compute (\ref{murph}) for $p=0.1$ only, and using the optimal
predictive, $P_{\boldsymbol{\hat{\theta}_{i}}}^{t-1}$, computed using CLS10
and QS10 respectively. In each case, the comparator (as made clear from the
notation used in (\ref{murph})) is the MLE-based predictive. Once again,
results are presented for all scenarios in Table \ref{tab:sim_design}.

\begin{figure}[tbp]
\centering
\includegraphics[width=500pt]{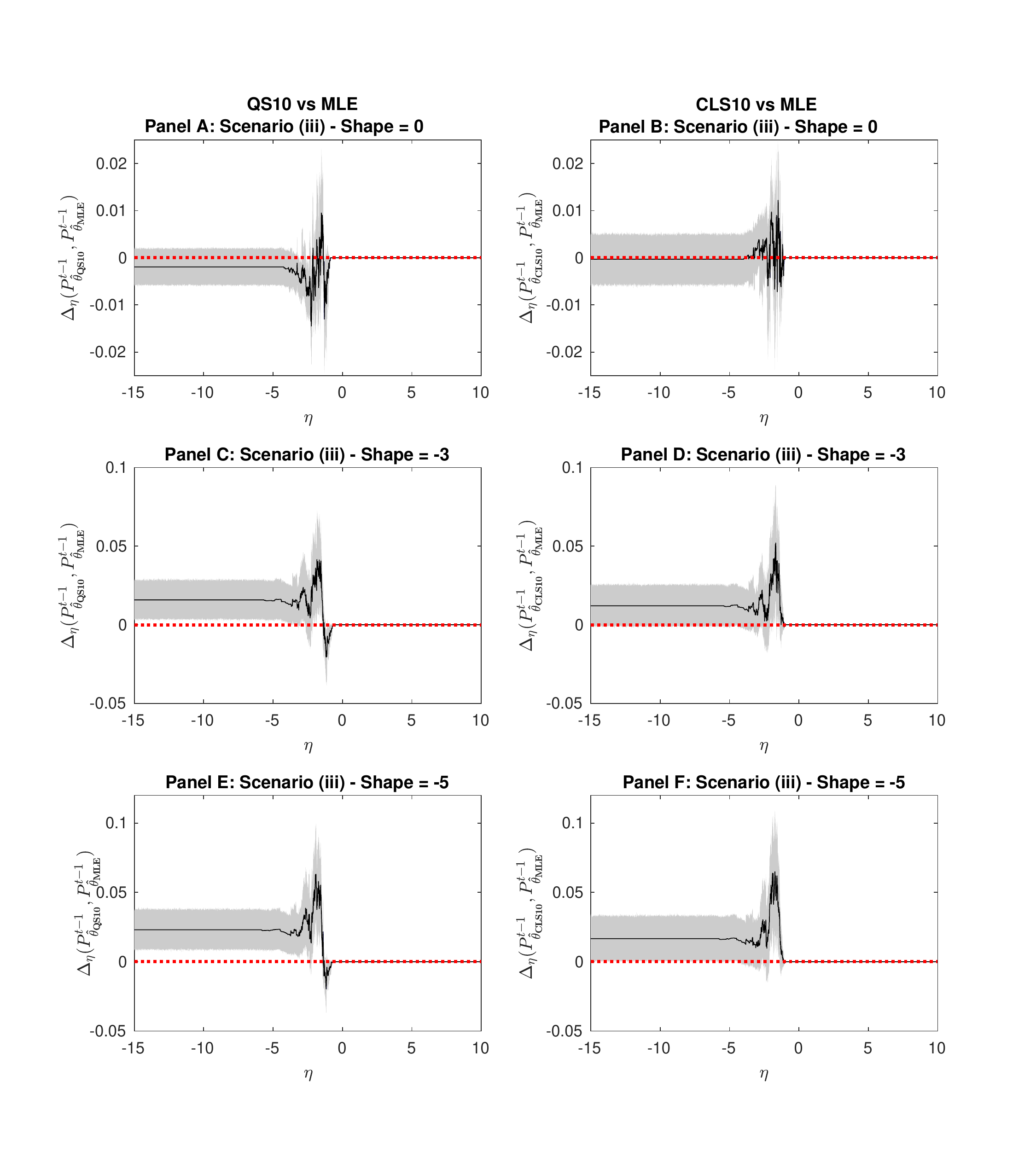}
\caption{10\% ES predictive performance for the DGPs in Scenario (iii) in
Table \protect\ref{tab:sim_design}. Each panel depicts the Murphy diagram,
plotting $\Delta _{\protect\eta }\left( P_{\boldsymbol{\hat{\protect\theta}%
_{i}}}^{t-1},P_{\boldsymbol{\hat{\protect\theta}_{MLE}}}^{t-1}\right) $
against $\protect\eta $, with the shaded area being the 95\% bootstrapped
confidence interval. The left hand side column assesses the QS10 predictive
relative to the MLE predictive, and the right hand side panel assesses the
CLS10 predictive relative to the MLE}
\label{fig:sv_sim_es}
\end{figure}

The Murphy diagrams for Scenarios (i) and (ii) are presented in Figure \ref%
{fig:garch_sim_es} and the diagrams for Scenario (iii) are depicted in
Figure \ref{fig:sv_sim_es}. The left hand column in each case depicts the
difference between the optimal QS10 predictive and the MLE predictive; while
the right hand column depicts the difference between the CLS10 predictive
and the MLE predictive. In each diagram, the average ES score difference is
presented as the solid black line, the bootstrapped 95\% confidence interval
is shaded in grey, and the benchmark of zero is the horizontal dotted red
line.

\begin{figure}[tbp]
\centering
\includegraphics[width=500pt]{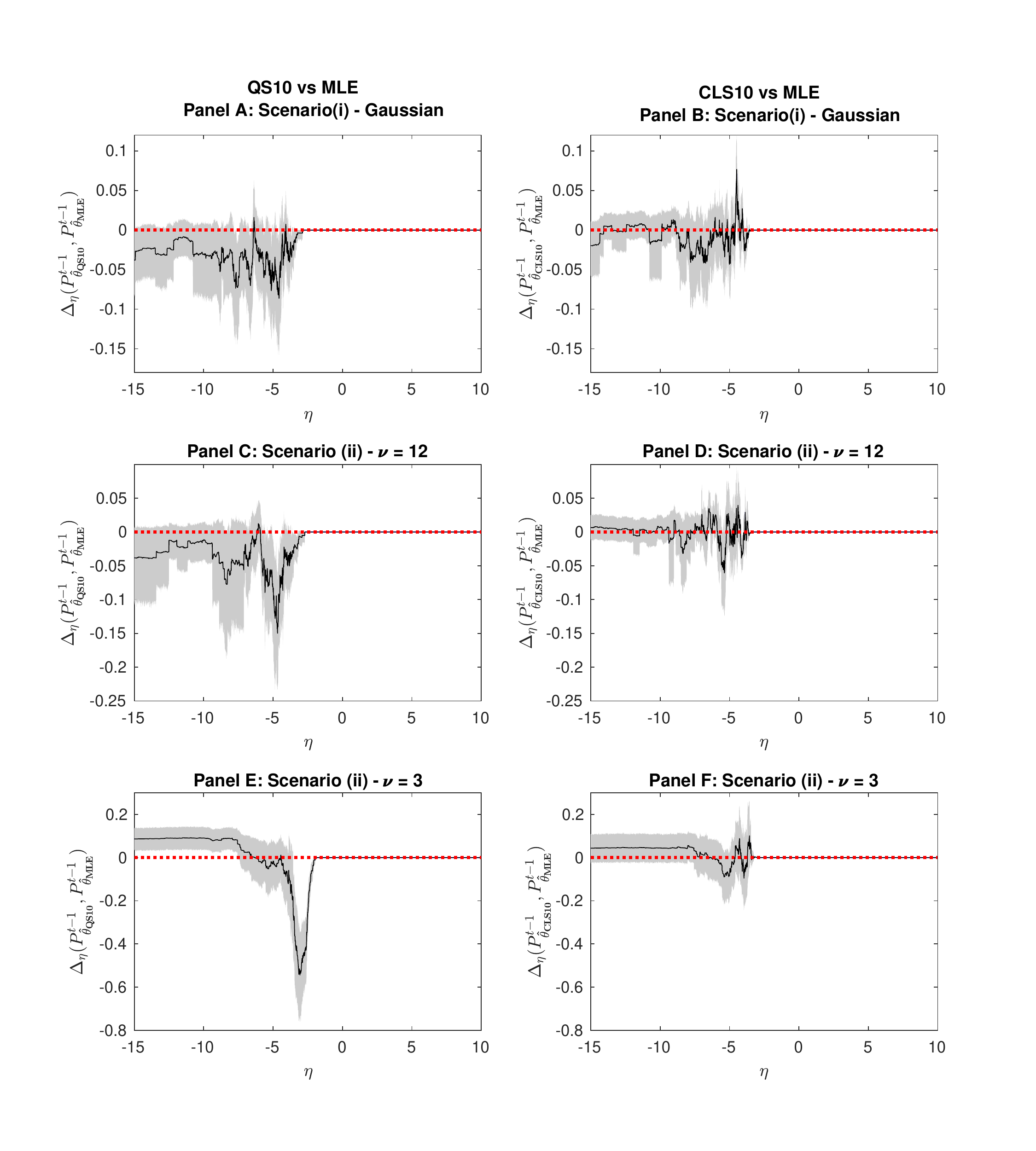}
\caption{10\% ES predictive performance for the DGPs in Scenario (i) and
(ii) in Table \protect\ref{tab:sim_design}. Each panel depicts the Murphy
diagram, plotting $\Delta _{\protect\eta }\left( P_{\boldsymbol{\hat{\protect%
\theta}_{i}}}^{t-1},P_{\boldsymbol{\hat{\protect\theta}_{MLE}}}^{t-1}\right) 
$ against $\protect\eta $, with the shaded area being the 95\% bootstrapped
confidence interval. The left hand side column assesses the QS10 predictive
relative to the MLE predictive, and the right hand side panel assesses the
CLS10 predictive relative to the MLE}
\label{fig:garch_sim_es}
\end{figure}

In Figure \ref{fig:garch_sim_es}, the QS10 optimal predictive dominates the
MLE predictive over a reasonable range of $\eta $ for the most misspecified
case in Scenario (ii), with $\nu =3$ (Panel E). The CLS10 predictive, on the
other hand, generates average scores that are larger in magnitude than those
of the MLE predictive, but the difference is not statistically significant
(Panel F). In the correctly specified and mildly misspecified cases (Panels
A - D), we do not observe a significant difference between either
tail-focused predictive and the MLE predictive.

Under Scenario (iii), we observe clearer benefits overall from using the
QS10 and CLS10 optimal predictives rather than the MLE predictive. As is to
be expected, when the true predictive is symmetric (Panels A and B), we do
not observe significant differences between the optimal and MLE average
scores. However, as the true predictive becomes more asymmetric (with
non-zero shape parameters), we observe a clear dominance of the QS10 and
CLS10 optimal predictives over the MLE predictive, as seen in Panels C - F
in Figure \ref{fig:sv_sim_es}. In addition, the average score differences
for the QS10 predictives are larger in magnitude than the corresponding
average score difference for the CLS10 predictives in these cases. This
suggests that calibrating the predictive distribution using the QS score
yields greater benefits in terms of accurate (joint) ES and VaR prediction
than using the CLS10 score. This conclusion is further supported by the
observation that the (greater) dominance (over the MLE predictive) of the
QS10 predictive relative to the CLS10 predictive coincides with the fact
that the former performs better overall than the latter in terms of the
VaR assessment in Section \ref{sec:simvar} (see Panel F, in particular, in Table \ref{tab:garch_var}).

\section{Empirical Analysis of Financial Returns}

\label{sec:returns}

In Section \ref{sec:sim}, using artificial data that mimic the features of
financial returns, we have illustrated the benefit of calibrating the
predictive distribution using the score that is appropriate for the
application of the prediction. In particular, we have observed that the
higher the degree of misspecification, the greater the benefit reaped by
optimal forecasting in this setting.

In this section, we produce optimal predictive distributions for returns on
a range of empirical financial indices: Standard and Poor's S\&P500, S\&P
Small Cap, S\&P Mid Cap and its nine industry indices, using the Gaussian
GARCH(1,1) model as the predictive model. Specifically, we document the
improvement in predictive performance that can be yielded by calibrating the
GARCH(1,1) model to scoring rules that focus predictive performance on the
sort of accuracy that matters in financial settings, despite the likely
misspecification of the predictive model itself. We adopt the following
scores in all of the empirical analysis, $S_{i}\in
\{LS,CSL10,CLS20,CLS80,CSL90,QS2.5,QS5,\linebreak QS10\}$. In Section \ref%
{pred} we assess the coherence of the predictive returns distributions to
the scoring rule used for calibration -- by matching the score used to
compute the out-of-sample average performance to the score used for
callibration. In Section \ref{pred} and \ref{es_emp} respectively we
document the improved accuracy of VaR and ES predictions that can be yielded
by the use of tail-focused scores.

\subsection{Data Description}

We consider the time period between 14 August 1996 to 30 July 2020 in our
empirical analysis. The data on the S\&P500 index are taken from Yahoo Finance, and data on all indices obtained from the Global Financial
Data (GFD) database. We construct 6000 continuously-compounded daily return
observations for each index, keeping the first 1000 observations for the
initial training sample and the latter 5000 observations for out-of-sample
predictive evaluations. Expanding estimation windows are used, as in the simulation exercise. The out-of-sample period between 1 August 2000 and 30
July 2020 includes two very volatile periods: the global financial crisis
(GFC) and the recent COVID-19 pandemic.

Table \ref{tab:spinx_stats} provides the details of the 12 indices we
investigate, and the descriptive statistics for all returns series. All
twelve return series are non-Gaussian, as evidenced by the rejections of the
Jarque-Bera test. As to be expected with daily financial returns, fat tails
are a prominent feature. We also observe negative skewness in all but the
S\&P45 index. The Ljung-Box test for serial correlation in the squared
returns is also indicative of time-varying volatility for all series.

\begin{table}[tbph]
\caption{Descriptions and summary statistics of the 12 S\&P indices under
investigation. Summary statistics include the mean, the standard deviation
(Std), and skewness and kurtosis statistics. The JB.Test column reports the
p-value of the Jarque-Bera test for normality and the LB.test column reports
the p-value of the test of serial correlation of the squared returns with 20
lags.}
\label{tab:spinx_stats}\centering
\resizebox{\textwidth}{!}{\begin{tabular}{llcccccc}
     & & & & & & & \\
    \toprule
   
    Index & Description & {Mean} & {Std } &{Skewness} & {Kurtosis} & {JB.Test} & {LB.Test} \\ \midrule
    S\&P500  & S\& 500 Market Index & 0.026 & 1.243 & -0.394 & 13.467 & 0.000 & 0.000 \\
    S\&P Small Cap & S\&P Small Capitalization Index & 0.031 & 1.244 & -0.553 & 10.938 & 0.000 & 0.000 \\
    S\&P Mid Cap & S\&P Mid Capitalization Index & 0.034 & 1.444 & -0.635 & 13.147 & 0.000 & 0.000 \\
    S\&P15 & S\&P Materials Index & 0.018 & 1.359 & -0.323 & 10.270 & 0.000 & 0.000 \\
    S\&P20 & S\&P Industrial Index & 0.023 & 1.527 & -0.419 & 10.857 & 0.000 & 0.000 \\
    S\&P25 & S\&P Consumer Discretionary Index & 0.035 & 1.373 & -0.274 & 11.194 & 0.000 & 0.000 \\
    S\&P30 & S\&P Consumer Staples Index & 0.025 & 1.378 & -0.175 & 13.683 & 0.000 & 0.000 \\
    S\&P35 & S\&P Health Care Index & 0.034 & 0.997 & -0.197 & 10.152 & 0.000 & 0.000 \\
    S\&P40 & S\&P Financial Index & 0.015 & 1.216 & -0.178 & 18.230 & 0.000 & 0.000 \\
    S\&P45 & S\&P Information Technology Index & 0.042 & 1.915 & 0.063 & 9.236 & 0.000 & 0.000 \\
    S\&P50 & S\&P Telecommunication Services Index & 0.007 & 1.772 & -0.044 & 9.941 & 0.000 & 0.000 \\
    S\&P55 & S\&P Utilities Index & 0.016 & 1.232 & -0.096 & 16.298 & 0.000 & 0.000 \\

    \bottomrule
    \end{tabular}}
\end{table}

\subsection{Performance of Optimal Prediction of Returns}

\label{pred}

Table \ref{tab:garch_inx} summarizes the out-of-sample predictive
performance of the GARCH(1,1) model for all 12 indices, and as measured
by all eight scoring rules. For each series, and in a particular column, we
report two numbers: i) the the average out-of-sample score for the MLE-based
predictive, and ii) the average out-of-sample score for the predictive that
is calibrated to the score used to measure performance, with this predictive
referred to as the ``optimal'' predictive. In the first column, for each
series, both of these numbers are automatically equivalent, given that the
score in question is the LS, and the MLE-based predictive is optimal
according to the LS. In all other columns, the numbers typically differ. The
bolded value for a given series, and in a given column, indicates the
predictive (MLE-based or optimal) that generates the higher average score
over the out-of-sample period, with an asterisk denoting rejection (at the
5\% significance level) of the null hypothesis of equal predictive ability
according to the test of \cite{GWtest2006}.

Overall, we observe predictive gain - and often significant predictive gain
- in calibrating the GARCH(1,1) model to the scoring rule of interest. In
particular, for all 12 series, there is significant predictive improvement
(as measured by both CLS10 and CLS20) produced by calibrating the
predictives to these same scoring rules. There is also improvement (and
sometimes significant improvement) in upper tail accuracy, to be had by
calibrating the predictives using CLS80 and CLS90. As made clear in earlier
work (see, e.g., \citealp{fbp2020}), such variation in relative performance
is not surprising, and is evidence of the fact that the misspecification of
the assumed predictive model impacts differently across the support of true
(unknown) predictive distribution which, itself, differs for each series.
That is, the assumed predictive model clearly performs poorly in the \textit{%
lower} tail (in particular) for all series, such that calibration according
to CLS10 and CLS20 reaps significant improvement. That said, the improvement
reaped by calibration via the lower tail quantile scores is not uniform
across series, although, as we will see in the following section, there are
certainly benefits to be had via these scoring rules when accurate
prediction of the VaR quantiles is the goal. Finally, we make the comment
that returns on the S\&P Mid Cap are the most skewed of all series
considered. This then coincides with a uniform significant improvement --
across all out-of-sample measures -- of the optimal predictive relative to
the MLE-based predictive. That is, misspecification of the symmetric
GARCH(1,1) model is most marked for this series, and the coherence of the
optimal predictions most in evidence as a consequence. It should also be
noted that scattered throughout the table there are a few instances where
the MLE predictive generates larger scores than the optimal predictive, but
in these instances the differences are statistically insignificant.

\begin{table}[tbph]
\caption{Average out-of-sample scores for returns on the 12 different S\&P
indices, with predictions produced using the Gaussian GARCH(1,1) model. For
each index, the model is calibrated using the LS (denoted by
``MLE''), as well as tail-focused score in the column heading (denoted by
``Optimal''). The column headings indicate the scores used to evaluate the
out-of-sample performance of the probabilistic forecasts..The bolded value
for a given series, and in a given column, indicates the predictive
(MLE-based or optimal) that generates the higher average score
over the out-of-sample period, with an asterisk denoting rejection (at the
5\% significance level) of the the null hypothesis of equal predictive
ability according to the test of \protect\cite{GWtest2006}.}
\label{tab:garch_inx}\centering
\resizebox{0.95\textwidth}{!}{\begin{tabular}{lcccccccc}

    &    &   &  & & & & &  \\
    \hline \hline
          & \multicolumn{8}{c}{Average out-of-sample scores} \\
\cmidrule{2-9}    Optimizers & LS   & CLS10 & CLS20 & CLS80 & CLS90  & QS2.5 & QS5   & QS10 \\
    \hline \hline
    \textbf{S\&P500} &            &       &       &       &       &       &       &  \\
    MLE    & -1.365  & -0.358 & -0.600 & -0.532 & -0.300 & -0.075 & -0.125 & -0.202 \\
    Optimal  & -1.365  & \textbf{-0.344${}^\star$} & \textbf{-0.584${}^\star$} & \textbf{-0.528${}^\star$} & \textbf{-0.296${}^\star$} &\textbf{ -0.074} & -0.125 & -0.202 \\
    \textbf{S\&P Small Cap} &              &       &       &       &       &       &       &  \\
    MLE    & -1.619  & -0.420 & -0.690 & -0.653 & -0.395 & -0.088 & -0.150 & -0.247 \\
    Optimal  & -1.619 & \textbf{-0.412${}^\star$} & \textbf{-0.682${}^\star$} & \textbf{-0.651} &\textbf{ -0.393} & \textbf{-0.086${}^\star$} & \textbf{-0.147${}^\star$} & \textbf{-0.245${}^\star$} \\
    \textbf{S\&P Mid Cap} &              &       &       &       &       &       &       &  \\
    MLE    & -1.493  & -0.382 & -0.649 & -0.577 & -0.346 & -0.082 & -0.138 & -0.227 \\
    Optimal  & -1.493  & \textbf{-0.372${}^\star$} & \textbf{-0.638${}^\star$} & \textbf{-0.572${}^\star$} & \textbf{-0.342${}^\star$} & \textbf{-0.079${}^\star$} & \textbf{-0.136${}^\star$} & \textbf{-0.225${}^\star$} \\
    \textbf{S\&P15} &              &       &       &       &       &       &       &  \\
    MLE    & -1.658  & -0.394 & -0.654 & -0.639 & -0.356 & -0.094 & -0.159 & -0.261 \\
    Optimal  & -1.658  & \textbf{-0.386${}^\star$} & \textbf{-0.645${}^\star$} & \textbf{-0.637} &\textbf{ -0.354 }& \textbf{-0.092} & \textbf{-0.158} & \textbf{-0.260} \\
    \textbf{S\&P20} &              &       &       &       &       &       &       &  \\
    MLE    & -1.521  & -0.401 & -0.643 & -0.565 & -0.339 & -0.087 & -0.144 & -0.232 \\
    Optimal  & -1.521  & \textbf{-0.386${}^\star$} & \textbf{-0.627${}^\star$} & \textbf{-0.563} & \textbf{-0.337${}^\star$} & \textbf{-0.084${}^\star$} & \textbf{-0.142} & \textbf{-0.231} \\
    \textbf{S\&P25} &              &       &       &       &       &       &       &  \\
    MLE    & -1.505  & -0.375 & -0.609 & -0.571 & -0.316 & -0.083 & -0.140 & -0.228 \\
    Optimal  & -1.505  & \textbf{-0.365${}^\star$} & \textbf{-0.597${}^\star$} & \textbf{-0.567${}^\star$} & \textbf{-0.315 }& \textbf{-0.081} & \textbf{-0.137${}^\star$} & -0.228 \\
    \textbf{S\&P30} &              &       &       &       &       &       &       &  \\
    MLE    & -1.157  & -0.298 & -0.506 & -0.474 &\textbf{ -0.251} & -0.058 & \textbf{-0.095} & -0.154 \\
    Optimal  & -1.157 & \textbf{-0.456${}^\star$} & \textbf{-0.285${}^\star$}  & -0.474 & -0.252 & -0.058 & -0.096 & -0.154 \\
    \textbf{S\&P35} &              &       &       &       &       &       &       &  \\
    MLE    & -1.367  & -0.317 & -0.554 & -0.491 & -0.237 & -0.072 & -0.121 & -0.195 \\
    Optimal  & -1.367  & \textbf{-0.308${}^\star$} & \textbf{-0.542${}^\star$} & \textbf{-0.489${}^\star$} & \textbf{-0.234${}^\star$} & -0.072 & -0.121 & -0.195 \\
    \textbf{S\&P40} &              &       &       &       &       &       &       &  \\
    MLE    & -1.668  & -0.403 & -0.630 & -0.588 & -0.351 & -0.107 & -0.176 & -0.283 \\
    Optimal  & -1.668  & \textbf{-0.393${}^\star$} & \textbf{-0.619${}^\star$} &\textbf{ -0.587} & -0.351 & \textbf{-0.105} & \textbf{-0.175} & -0.283 \\
    \textbf{S\&P45} &              &       &       &       &       &       &       &  \\
    MLE    & -1.686  & -0.291 & -0.547 &\textbf{ -0.507} &\textbf{ -0.250} & -0.097 & -0.166 & -0.273 \\
    Optimal  & -1.686  & \textbf{-0.282${}^\star$} & \textbf{-0.536${}^\star$} & -0.508 & -0.252 & -0.097 & -0.166 & -0.273 \\
    \textbf{S\&P50} &              &       &       &       &       &       &       &  \\
    MLE    & -1.566  & -0.340 & -0.597 & -0.543 & -0.299 & -0.089 & -0.147 & -0.235 \\
    Optimal  & -1.566  & \textbf{-0.329${}^\star$} & \textbf{-0.586${}^\star$} & \textbf{-0.542} & -0.299 &\textbf{ -0.088 }& -0.147 & -0.235 \\
    \textbf{S\&P55} &              &       &       &       &       &       &       &  \\
    MLE    & -1.419  & -0.390 & -0.651 & -0.574 & -0.333 & -0.078 & -0.130 & \textbf{-0.210} \\
    Optimal  & -1.419 & \textbf{-0.381${}^\star$} & \textbf{-0.641${}^\star$} & \textbf{-0.572${}^\star$} & \textbf{-0.330${}^\star$} & -0.078 & -0.130 & -0.212 \\
    \hline \hline
    \end{tabular}}
\end{table}

\subsection{Value-at-Risk Prediction\label{var}}

\label{sec:VaR}

We now document the accuracy of VaR predictions based on the MLE and optimal
(QS) predictives. The optimal QS refers to the predictive calibrated via the
QS score in (\ref{eq:qs}), with $p\times 100=$ $2.5$, $5$ and $10$ used
respectively for prediction of the VaR at the $2.5\%$, $5\%$ and $10\%$
nominal levels. Empirical coverages are recorded in Table \ref{tab:inx_var},
with bolded values indicating the instances where the null hypothesis of the
conditional coverage test of \cite{christoffersen1998} is not rejected at
the 5\% significance level.

Consistent with the observations from the simulation exercise in Section \ref%
{sec:simvar}, calibrating the VaR prediction using the appropriate QS score
leads to better coverage results overall. For all but one of the 12
indices, the optimal QS prediction produces 2.5\% VaR coverages that are not
statistically different from the nominal coverage level, but only two of the
12 MLE predictions pass the conditional coverage test. For the 5\% VaR,
the optimal QS prediction produces appropriate coverage for all indices,
while the MLE prediction produces appropriate coverage for only seven
indices. In the case of the 10\% VaR, both the MLE and optimal QS perform
comparably, each producing appropriate coverages for half of the indices
considered. This result shows that despite the optimal QS predictive
performing on a par with the MLE predictive in the QS evaluation columns in
Table \ref{tab:garch_inx}, the optimal QS predictive -- when used
specifically to accurately predict the corresponding VaR quantile -- does
produce notably better outcomes overall that the MLE-based predictive,
particularly for the lower values of $p.$

\begin{table}[tbph]
\caption{Predictive VaR coverage statistics, based on the Gaussian
GARCH(1,1) predictive model, for the 12 index return series. The ``Optimal
QS'' refers to the predictive distribution calibrated using the QS score for $%
p\times 100=$ $2.5$, $5$ and $10.$ Bolded values indicate the instances
where the null hypothesis of the conditional coverage test of \protect\cite%
{christoffersen1998} is not rejected at the 5\% significance level.}
\label{tab:inx_var}\centering
\begin{tabular}{llccc}
&  &  &  &  \\ \hline\hline
&  & \multicolumn{3}{c}{Out-of-sample coverage} \\ 
\cmidrule{3-5} Index & Optimizers & 2.5\% VaR & 5\% VaR & 10\% VaR \\ 
\hline\hline
{S\&P 500} & MLE & 3.400\% & \textbf{5.300\%} & \textbf{10.040\%} \\ 
& Optimal QS & \textbf{2.540\%} & \textbf{4.480\%} & 8.980\% \\ 
\cmidrule{2-5} {S\&P Small Cap} & MLE & 3.660\% & 6.420\% & 11.480\% \\ 
& Optimal QS & 3.160\% & \textbf{5.400\%} & 10.800\% \\ 
\cmidrule{2-5} {S\&P Mid Cap} & MLE & 3.540\% & 6.240\% & 11.120\% \\ 
& Optimal QS & \textbf{2.800\%} & \textbf{4.920\%} & \textbf{9.820\%} \\ 
\cmidrule{2-5} {S\&P 15} & MLE & 3.280\% & 5.640\% & \textbf{9.780\%} \\ 
& Optimal QS & \textbf{2.840\%} & \textbf{5.540\%} & \textbf{10.200\%} \\ 
\cmidrule{2-5} {S\&P 20} & MLE & 3.560\% & \textbf{5.280\%} & \textbf{9.900\%%
} \\ 
& Optimal QS & \textbf{2.580\%} & \textbf{4.960\%} & \textbf{10.080\%} \\ 
\cmidrule{2-5} {S\&P 25} & MLE & \textbf{3.220\%} & \textbf{5.420\%} & 
\textbf{9.880\%} \\ 
& Optimal QS & \textbf{2.600\%} & \textbf{5.160\%} & \textbf{9.520\%} \\ 
\cmidrule{2-5} {S\&P 30} & MLE & \textbf{2.640\%} & \textbf{4.840\%} & 
8.660\% \\ 
& Optimal QS & \textbf{2.400\%} & \textbf{4.640\%} & 8.620\% \\ 
\cmidrule{2-5} {S\&P 35} & MLE & 3.100\% & \textbf{5.000\%} & 8.480\% \\ 
& Optimal QS & \textbf{2.380\%} & \textbf{4.440\%} & 8.880\% \\ 
\cmidrule{2-5} {S\&P 40} & MLE & 3.060\% & \textbf{5.260\%} & \textbf{9.100\%%
} \\ 
& Optimal QS & \textbf{2.540\%} & \textbf{5.000\%} & \textbf{9.240\%} \\ 
\cmidrule{2-5} {S\&P 45} & MLE & 3.360\% & 5.920\% & \textbf{9.980\%} \\ 
& Optimal QS & \textbf{2.520\%} & \textbf{4.840\%} & \textbf{9.220\%} \\ 
\cmidrule{2-5} {S\&P 50} & MLE & 3.240\% & \textbf{5.240\%} & 9.040\% \\ 
& Optimal QS & \textbf{2.520\%} & \textbf{5.020\%} & \textbf{9.420\%} \\ 
\cmidrule{2-5} {S\&P 55} & MLE & 3.520\% & 5.660\% & 10.020\% \\ 
& Optimal QS & \textbf{2.840\%} & \textbf{5.380\%} & 10.320\% \\ \hline\hline
\end{tabular}%
\end{table}

\subsection{Expected Shortfall Prediction\label{es_emp}}

\begin{figure}[tbp]
\centering
\includegraphics[width=500pt]{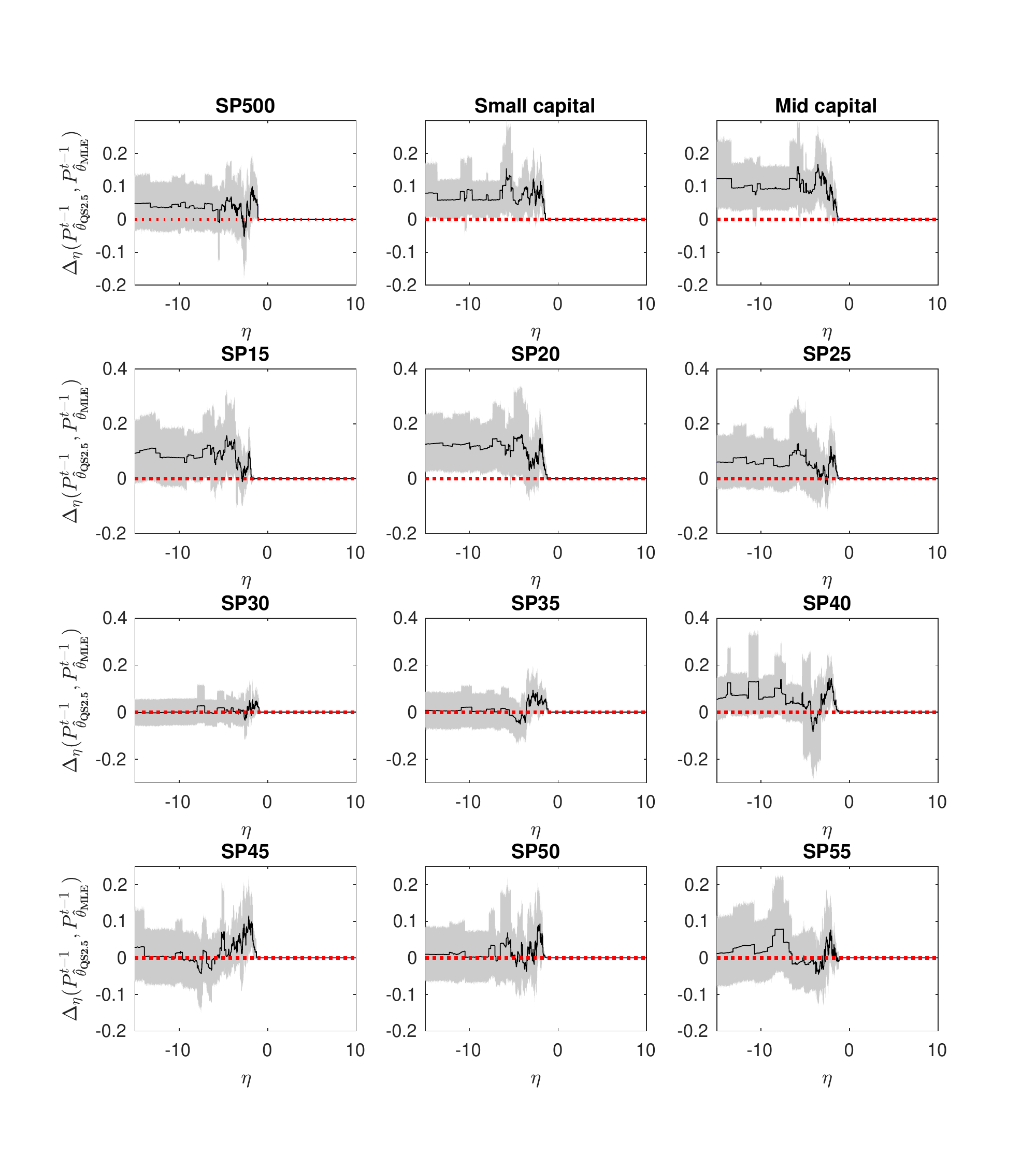}
\caption{2.5\% ES performance for the 12 S\&P indices using the Gaussian
GARCH(1,1) as the predictive model. All Murphy diagrams depict $\Delta _{%
\protect\eta }\left( P_{\boldsymbol{\hat{\protect\theta}_{QS2.5}}%
}^{t-1},P_{\boldsymbol{\hat{\protect\theta}_{MLE}}}^{t-1}\right) $, with the
shaded area being the 95\% bootstrapped confidence interval.}
\label{fig:QS25_ES}
\end{figure}

In Section \ref{sec:es_sim}, we established that when the predictive
distribution calibrated using the QS score outperformed the MLE predictive
in terms of VaR prediction, we also observed dominance of the QS predictive
in terms of ES prediction. In this section, we verify this observation
empirically by assessing the ES predictions generated from the optimal QS
predictive and the MLE predictive. Figures \ref{fig:QS25_ES}-\ref%
{fig:QS10_ES} present the Murphy diagrams for the 2.5\%, 5\% and 10\% ES,
respectively, for the 12 return series. We reiterate that if the bootstrap
confidence interval is above zero for a reasonable range of values for $\eta 
$, the QS predictive is deemed to outperform the MLE predictive in ES
prediction.

\begin{figure}[tbp]
\centering
\includegraphics[width=500pt]{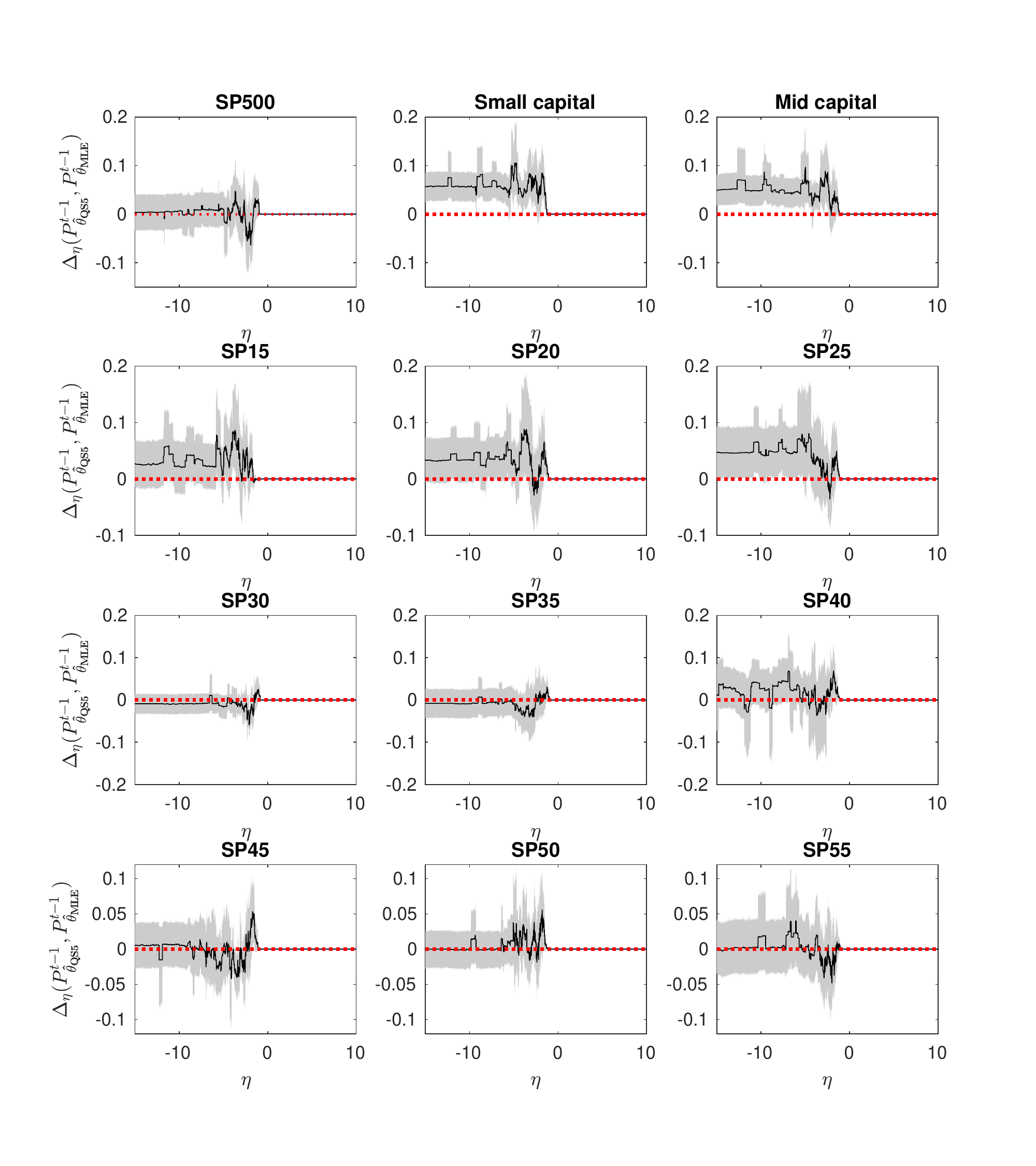}
\caption{5\% ES performance for the 12 S\&P indices using the Gaussian
GARCH(1,1) as the predictive model. All Murphy diagrams depict $\Delta _{%
\protect\eta }\left( P_{\boldsymbol{\hat{\protect\theta}_{QS5}}}^{t-1},P_{%
\boldsymbol{\hat{\protect\theta}_{MLE}}}^{t-1}\right) $, with the shaded
area being the 95\% bootstrapped confidence interval.}
\label{fig:QS5_ES}
\end{figure}

In Figure \ref{fig:QS25_ES}, the average score difference is statistically
greater than zero for the 2.5\% ES prediction for the S\&P Small Cap, S\&P
Mid Cap and S\&P20 indices. Dominance of the optimal QS predictive is
observed for the S\&P Small Cap, S\&P Mid Cap and S\&P25 indices for the
case of the 5\% ES (Figure \ref{fig:QS5_ES}), and for the 10\% ES dominance is
observed for the S\&P Small Cap and S\&P Mid Cap indices (Figure \ref%
{fig:QS10_ES}). These results are consistent with the almost uniform superior
performance of the QS predictives (over the MLE predictives) in Table \ref{tab:garch_inx} for the
corresponding sets of series.

\begin{figure}[tbp]
\centering
\includegraphics[width=500pt]{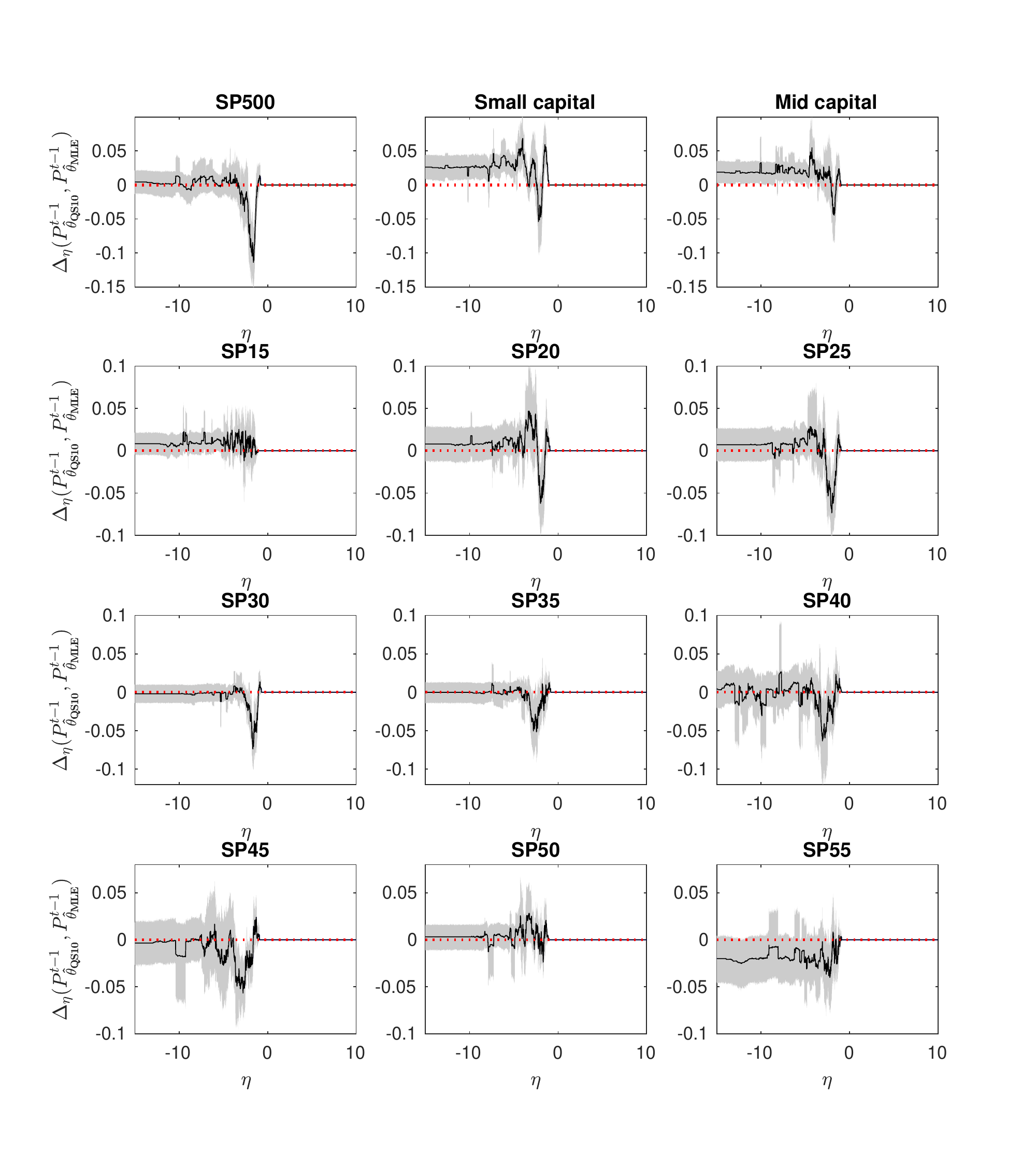}
\caption{10\% ES performance for the 12 S\&P indices using the Gaussian
GARCH(1,1) as the predictive model. All Murphy diagrams depict $\Delta_{%
\protect\eta}\left(P^{t-1}_{\boldsymbol{\hat{ \protect\theta}_{QS10}}%
},P^{t-1}_{\boldsymbol{\hat{ \protect\theta}_{MLE}}}\right)$, with the
shaded area being the 95\% bootstrapped confidence interval.}
\label{fig:QS10_ES}
\end{figure}

\section{Optimizing the Predictive VIX Distribution for Dynamic Risk
Management}

\label{sec:vix}

The VIX is a volatility index published by the Chicago Board of Exchange
(CBOE), and is a measure of US market volatility constructed from option
prices. It is well known that the VIX index is negatively correlated with
the stock market index, with the VIX typically rising in periods of market
stress when there is a general decline in the stock market index. Such
negative correlation is also observed between the stock index and the
tradable futures contract written on the VIX index, with the correlation
reportedly being stronger in the futures market \citep{moran2020vix}. As
such, VIX futures contracts are often used as a tool to hedge against
market downturns. In this section, we illustrate the benefits of optimizing
the predictive distribution of the VIX index for the purpose of improving risk
management strategies that involve a position in VIX futures.

\subsection{The Predictive Model for the VIX index}

The VIX index is a model-free measure of option implied volatility, extracted from options written on the S\&P500 stock
market index (see \citealp{exchange2009cboe}, for more details). In theory,
the VIX index measures the risk-neutral forward-looking expectation of the
integrated variance of the underlying asset (see \citealp{britten2000option}%
, and \citealp{jiang2005}). Thus, the movement in the VIX index not only
closely tracks volatility itself, but also contains information about the
forward-looking expectations of derivative traders, and is often viewed as a
leading indicator of market movements as a result.

We have collected daily VIX index data from Yahoo Finance for the period: 27th
September 1996 to 30th July 2020. The descriptive statistics of $\log
(VIX_{t})$ are shown in Table \ref{tab:vixstats}. The statistics indicate
that the marginal density of $\log (VIX_{t}) $ is positively skewed and far
from being normally distributed. As is typical with volatility measures, the
VIX index has strong persistence, with its autocorrelation function
remaining high in magnitude and statistically significant as the lag
increases (with 20 lags being used for the Ljung-Box test of autocorrelation
in Table \ref{tab:vixstats}).

\begin{table}[tbp]
\caption{Descriptive statistics for the logarithm of the VIX volatility
index. The Min and Max are the smallest and largest value of the sample
series, respectively. Skewness is a measure of the asymmetry and kurtosis is
a measure of fat tails with the normal distribution having a kurtosis equal
to three. `JB.Test' is the test statistic for the Jarque-Bera test of
normality and the p-value is recorded in the table. `LB.Test' is the test of
serial correlation of the squared volatility based on 20 lags and the
p-value is recorded in the table.}
\label{tab:vixstats}\centering
\begin{tabular}{lrrrrrrrr}
&  &  &  &  &  &  &  &  \\ 
\cmidrule(l{3pt}r{3pt}){1-9} Series & Min & Median & Mean & Max & Skewness & 
Kurtosis & JB.Test & LB.Test \\ 
\midrule Log(VIX) & 2.213 & 2.916 & 2.936 & 4.415 & 0.604 & 3.363 & 0.000 & 
0.000 \\ 
\bottomrule &  &  &  &  &  &  &  & 
\end{tabular}%
\end{table}

In order to capture the long memory feature of the VIX index, we employ the
heterogeneous autoregressive (HAR) model to construct the predictive
distribution of the VIX index. Proposed by \citet{corsi2009}, the HAR model
is a simple additive linear model that uses moving averages of lagged values
of the relevant dependent variable as regressors. It does not belong to the
class of long memory models, but it is able to produce persistence that is
almost indistinguishable from what is observed in financial volatility in
practice, through the simple autoregressive-type structure. The HAR model
and its extensions have been employed by previous studies on the
predictability of the VIX index; see, for example, \citet{manee2012probabilistic} and \citet{fernandes2014}. We
include a GARCH component in the HAR model to account for time variation in
the second moment of the VIX distribution. The HAR-GARCH model is defined as:

\begin{equation}
\log (VIX_{t+1})=\beta _{0}+\beta _{1}\log (VIX_{t})+\beta _{2}\log
(VIX_{t-5,t})+\beta _{3}\log (VIX_{t-22,t})+\sigma _{t+1}z_{t+1}
\label{eq:HAR}
\end{equation}%
where 
\begin{equation}
\sigma _{t+1}^{2}=\alpha _{0}+\alpha _{1}(\sigma _{t}z_{t})^{2}+\alpha
_{2}\sigma _{t}^{2},  \label{eq:garch}
\end{equation}%
\begin{equation}
z_{t+1}\sim i.i.d.N(0,1)
\end{equation}%
and 
\begin{equation}
\log (VIX_{t-m,t})=\frac{1}{m}\sum_{i=1}^{m}\log (VIX_{t+i-1}).
\label{eq:ma5}
\end{equation}

\subsection{Performance of Optimal Prediction of the VIX Index}

Using a similar expanding window predictive assessment procedure to that conducted in
Sections \ref{sec:sim_design} and \ref{pred}, with an initial sample size equal to 1000 and
out-of-sample size $\tau =T-1000=5000$, we produce and evaluate predictive
densities based on the HAR-GARCH as defined in (\ref{eq:HAR}) to (\ref%
{eq:ma5}). We optimize the predictive distribution using the LS and the CLS
scores corresponding to the 10\% and 20\% lower-tail, as well as the 80\%
and the 90\% upper-tail of the marginal VIX distribution. Evaluation of
out-of-sample performance uses this same set of scores.

Table \ref{vix_coherence} documents the average out-of-sample scores for the
MLE-based and optimal predictives. Once again, bolded values represent the
larger out-of-sample average score in a given column, with an asterisk
denoting a case where the difference between the MLE and optimal predictive
scores is significantly different from zero according to the test of %
\citet{GWtest2006}. The results from this table show that the optimal
forecasts generate larger out-of-sample average scores in \textit{all} cases (other than when the MLE and optimal predictives coincide), with
the improvement statistically significant in the case of the CLS10
predictive.

\begin{table}[tbp]
\caption{Average out-of-sample scores for prediction of the VIX index where
the predictive model is the HAR-GARCH model with a Gaussian error. The model
is calibrated using the LS (denoted by ``MLE''), as well as the CLS in the column
heading (denoted by ``Optimal''). The column headings indicate the scores used
to evaluate the out-of-sample performance of the probabilistic forecasts. The bolded value in a given column indicates the predictive (MLE-based or
optimal) that generates the higher average score over the out-of-sample
period, with an asterisk denoting rejection (at the 5\% significance level)
of the the null hypothesis of equal predictive ability according to the test
of \protect\cite{GWtest2006}.}
\label{vix_coherence}\centering
\begin{tabular}{lcccccc}
&  &  &  &  &  &  \\ \hline\hline
&  &  &  &  &  &  \\ 
Optimizers & \multicolumn{6}{c}{Average out-of-sample scores} \\ \hline\hline
& LS & CLS10 & CLS20 & CLS80 & CLS90 \\ \cline{2-7}
MLE & 1.302 & 0.344 & 0.566 & 0.131 & 0.059 \\ 
Optimal & 1.302 & \textbf{0.350*} & \textbf{0.569} & 
\textbf{0.134} & \textbf{\ 0.061} \\ \hline\hline
\end{tabular}%
\end{table}

\subsection{Application to Dynamic Risk Management with VIX Futures}

\citet{moran2020vix} demonstrate how VIX futures can be used for dynamic
risk management. They outline a backtesting strategy that involves the
static allocation of a 95\% long position in the stock portfolio and a 5\% long position in VIX futures. Since the VIX
futures derive their value from the spot VIX index, they also propose a
dynamic allocation strategy where the position in the VIX futures depends on
the level of the spot VIX index itself.

To illustrate the possible benefit of score-based calibration in such a
setting, we adapt the dynamic strategy illustrated in \citet{moran2020vix}
to incorporate the predictive distribution for the VIX index based on the
HAR-GARCH model, but optimized using an appropriate version of the
tail-focused CLS. We design two dynamic trading strategies, in which the
position on the shortest maturity VIX future is triggered by one of two
decision rules based on the optimal predictive for the VIX
index.\footnote{%
We note related studies by \citet{konstantinidi2008} and %
\citet{taylor2019} in which trading strategies related to VIX futures, 
based on point predictions, are constructed.} The opening and closing prices of the VIX
futures contracts over the period between 24 September 2008 and 28 May 2021
are obtained directly from the CBOE website.

First, we backtest a predictive probability rule, where the 5\% long
position in the VIX futures is taken only if $\text{Pr}(VIX_{t+1}>VIX_{t}|%
\mathcal{F}_{t})>0.5$. That is, when the VIX index is likely to increase
over the next day, we take a long position in the VIX futures in order to
offset the potential fall that is expected in the stock market. To implement
the predictive probability rule, we optimize the VIX predictive distribution
according to the upper-tail CLS associated with the region where $%
VIX_{t}>VIX_{t-1}$, denoted by CLS($VIX_{t}>VIX_{t-1}$).

Second, we utilize the $80^{th}$ predictive VIX percentile to form a trading
rule. Under this rule, the 5\% long position in the VIX futures market is
only undertaken if the $80^{th}$ predictive percentile of the spot VIX is
greater than 40\%. That is, the value that defines the upper 20\% of the
predictive distribution of the VIX must be high enough for the investor to
take a position to offset the potential fall in the stock market. To
implement this rule, the predictive distribution of the VIX is optimized
according to CLS80.

Under both dynamic settings, the decision to adopt the VIX futures position
is made at the beginning of the trading day, based on the relevant predictive
distribution constructed from data up to the previous
trading day. If a decision is made to enter a position in the VIX futures, a
long position in the shortest maturity VIX futures contract is taken. The
VIX futures\textbf{\ }position is closed off at day end, with the investor
realizing the return that reflects the change from the opening price to the
closing price of the VIX futures on that trading day.

Table \ref{tab:VIXtrade} reports portfolio performance measures based on
each of the dynamic strategies outlined above. The performance of the static
portfolio, comprising a 95\% long position in the S\&P500 market portfolio
and a 5\% long position in the shortest maturity VIX futures, as in \cite%
{moran2020vix}, is also reported for comparison. For each dynamic strategy,
we also produce results using the predictive distribution based on the MLE, for comparison. We
report the mean excess return, that is, the portfolio return over and above
the risk-free rate of return, the portfolio standard deviation, as well as
the Sharpe ratio. 

Under both dynamic trading rules, the predictive distributions
constructed from the upper-tail CLS generate notably larger Sharpe ratios than do the MLE predictives,
implying that the strategy undertaken with upper-tail calibration generates
the greatest return relative to the risk undertaken. This observation
confirms the need -- in an empirical setting like this, in which the model misspecification prevails -- to construct the predictive distribution that is optimal
to the region of interest, in this case in the upper tail of the predictive
VIX distribution. We also observe that the
predictive percentile rule performs better than the predictive
probability rule -- for both forms of predictive. In fact, the portfolio constructed using the predictive
probability rule and exploiting the MLE-based
predictive performs worse than the static portfolio. This highlights the importance of region-specific
calibration in rendering the dynamic risk management
strategies unambiguously preferable to the simpler static option.

\begin{table}[htbp]
\caption{Stock and VIX futures portfolio performance based on dynamic
trading strategies. The results for the static portfolio made up of 95\%
S\&P500 market portfolio and 5\% VIX futures on all days, are also reported
for comparison.}
\label{tab:VIXtrade}\centering
\begin{tabular}{llccc}
\toprule\toprule &  &  &  &  \\ 
& {Mean Ex. Returns} & {Std. Dev.} & {Sharpe Ratio} &  \\ \cline{2-4}
&  &  &  &  \\ 
Static Portfolio & 0.040 & 0.182 & 0.219 &  \\ \hline
&  &  &  &  \\ 
& \multicolumn{3}{c}{\textbf{\ Predictive probability rule}} &  \\ 
Optimizer &  &  &  &  \\ 
CLS($VIX_{t}>VIX_{t-1}$) & 0.066 & 0.207 & {\textbf{0.318}} &  \\ 

MLE & 0.042 & 0.204 & 0.206 &  \\ \hline
&  &  &  &  \\ 
& \multicolumn{3}{c}{\textbf{\ $80^{th}$ Predictive percentile rule}} &  \\ 
Optimizer &  &  &  &  \\ 
CLS80 & 0.076 & 0.199 & {\textbf{0.381}} &  \\ 

MLE & 0.075 & 0.200 & 0.375 &  \\ \hline\hline
\end{tabular}%
\end{table}

\section{Conclusion}

\label{sec:conclude}

In this paper, we investigate the use of tail-focused optimal forecasts in
the context of financial risk management. From both the simulation and
empirical results, we establish the benefits of optimal forecasts in
misspecified models, subject to the assumed predictive model being broadly
suitable for modelling the data at hand. Focusing on the tail
performance of optimal forecasts, the simulation results show that the more
misspecified the model is, the more benefits can be gained from using the
optimal forecasts, in comparison with adopting the conventional MLE-based
forecasts. In particular, the gain from using optimal forecasts is more
stark in the case of a skewed DGP, than in the case of fat-tailed DGP, when
the assumed predictive model is conditionally Gaussian. We also establish
that when tail risk is of interest, the optimal forecasts constructed from
the quantile score yield better predictions than other alternatives in the
context of value-at-risk coverage and expected shortfall accuracy.

The empirical analysis of 12 S\&P indices confirms the qualitative
nature of our simulation results. We also investigate the use of optimal
forecasting in the context of VIX prediction, with a specific application to
hedging stock market risk with VIX derivatives. Trading strategies
constructed from tail-focused optimal forecasts are backtested, alongside
trading strategies constructed from MLE-based forecasts, with the
tail-focused forecasts generating substantially better return-risk tradeoffs in this
context.

\medskip \baselineskip14pt 
\bibliographystyle{Myapalike}
\bibliography{references}

\end{document}